\documentclass{amsart}

\title{A Clifford algebra model in phase space}
\author{Robert Arnott Wilson}
\date{1st April 2024}

\address{Queen Mary University of London}
\email{r.a.wilson@qmul.ac.uk}

\newcommand{\CCl}{\mathbb C\ell}
\newcommand{\Cl}{\mathcal C\ell}
\newcommand{\CC}{\mathbb C}

\newcommand{\RR}{\mathbb R}

\newcommand{\HH}{\mathbb H}

\begin{document}
\begin{abstract}
I show how the isomorphism between the Lie groups of types $B_2$ and $C_2$ leads to a faithful action of the Clifford algebra $\mathcal C\ell(3,2)$ on the phase space of
$2$-dimensional dynamics, and hence to a mapping from Dirac spinors modulo scalars into this same phase space. Extending to the phase space of $3$-dimensional
dynamics allows one to embed all the gauge groups of the Standard Model as well, and hence unify the electro-weak and strong forces into a single algebraic structure,
identified as the gauge group of Hamiltonian dynamics. The gauge group transforms between phase space coordinates appropriate for arbitrary observers, and therefore
shows how the apparently arbitrary parameters of the Standard Model transform between mutually accelerating observers. In particular, it is possible to calculate the
transformation between an inertial frame and the laboratory frame, in order to explain how macroscopic laboratory mechanics emerges from quantum mechanics,
and to show how to write down a quantum theory of gravity that is consistent with quantum mechanics, but is not consistent with General Relativity. 
\end{abstract}
\maketitle

\section{Electo-weak theory}
\subsection{Clifford algebras and spin groups}
The Dirac algebra \cite{Dirac}, that is central to the Standard Model of Particle Physics (SMPP), is a copy of the algebra $M_4(\CC)$ of $4\times 4$
complex matrices, that is usually described, or interpreted, as the complex Clifford algebra $\CCl(3,1)$ of Minkowski spacetime \cite{BjorkenDrell,Griffiths,WoitQFT}.
However, Clifford algebras are naturally real, not complex, and the fundamental reason for the complexification is not readily apparent,
although it is clearly necessary for practical calculations.
In fact, there are three real Clifford algebras isomorphic to $M_4(\CC)$, namely
$\Cl(4,1)$, $\Cl(2,3)$ and $\Cl(0,5)$.  The first two are often interpreted to extend Minkowski spacetime to include an extra dimension of either time or space, resulting in
what is called de Sitter space or anti-de Sitter space respectively.

The main purpose of constructing a Clifford algebra is to construct a spin group \cite{Zee}, and the Dirac algebra allows us to construct all three real forms of $Spin(5)$,
\begin{itemize}
\item $Spin_5(\RR)\cong Sp_2(\HH)$;
\item $Spin_{4,1}(\RR)\cong Sp_{2,2}(\RR)$;
\item $Spin_{3,2}(\RR)\cong Sp_4(\RR) \cong Sp_2(\HH')$.
\end{itemize}
The notations for these groups are not well standardised, and many different forms of notation are used in the literature. 
Since we shall only be interested in the third case, there should be no
confusion if we denote it $Sp(2)$, as is common in the physics literature on phase space, even though this is not strictly correct. The more specific notation $Sp_2(\HH')$
indicates that its Lie algebra consists of anti-Hermitian matrices over the split form of the quaternions. Here we use the 
definition 
$[A,B]:=AB-BA$ for the Lie bracket, without an extra factor of $i$. 

\subsection{Pauli and Dirac matrices}
Recall that the compact form $\HH$ of the quaternions is defined by Hamilton's famous equations
\begin{align}
i^2=j^2=k^2=ijk=-1.
\end{align}
The split form $\HH'$ can be defined by moving the negative sign to the left:
\begin{align}
-I^2=J^2=K^2=IJK=1.
\end{align}
The Hermitian $2\times 2$ matrices over $\HH'$  
correspond to the Dirac $\gamma$ matrices as follows:
\begin{align}
&\sigma^1 := \begin{pmatrix}1&0\cr 0&-1\end{pmatrix}\mapsto i\gamma^1,\quad
\sigma^2: = \begin{pmatrix}0&1\cr 1 & 0\end{pmatrix}\mapsto i\gamma^2,\quad
\sigma^3 := \begin{pmatrix}0&-I\cr I & 0\end{pmatrix}\mapsto i\gamma^3,\cr
&\qquad \sigma^4 := \begin{pmatrix}0&-J\cr J & 0\end{pmatrix}\mapsto i\gamma^0,\quad
\sigma^5 := \begin{pmatrix}0&-K\cr K & 0\end{pmatrix}\mapsto i\gamma^5.
\end{align}
The notation is chosen to show how the Dirac matrices are a natural generalisation of the Pauli matrices, when we extend from $\CC$ to $\HH'$.
The products of pairs of these matrices are the anti-Hermitian generators for the Lie algebra $\mathfrak{sp}_2(\HH')\cong \mathfrak{so}(3,2)$.

The Dirac spinors on which these matrices act can be written as columns of two split quaternions, and therefore have 8 real degrees of freedom, as they 
do in standard quantum mechanics.
There is a complex structure defined by multiplication by the complex scalar
\begin{align}
\sigma^1\sigma^2\sigma^3 = \begin{pmatrix}I&0\cr 0&I\end{pmatrix}
\end{align}
and the equations 
\begin{align}
IJ=K,\quad  IK=-J
\end{align}
show that if the $1,I$ coordinates of the spinor are taken to represent the left-handed Weyl spinor, then the $J,K$ coordinates represent the right-handed Weyl spinor, and vice versa.

\subsection{The Jordan algebra}
The fundamental splitting of the matrix algebra into Hermitian and anti-Hermitian parts is a splitting into a Jordan algebra (Hermitian matrices) under the Poisson bracket
\begin{align}
\{A,B\}:=AB+BA
\end{align}
and a Lie algebra (anti-Hermitian matrices) under the Lie bracket
\begin{align}
[A,B]:=AB-BA.
\end{align}
In physics, Jordan algebras model fermions and Lie algebras model bosons.
The Jordan algebra has an identity element
\begin{align}
\begin{pmatrix}1&0\cr 0&1\end{pmatrix} = -\sigma^1\sigma^2\sigma^3\sigma^4\sigma^5,
\end{align}
which is used in the Dirac equation for the mass of a fermion (multiplied by $i$ to convert from our convention to the standard convention).
The five Dirac matrices all square to $\pm1$, and split into a set of three with positive mass, and a completely different set of two with negative mass.
Since particles with negative mass cannot exist as free particles, these must be interpreted as quarks, which cannot exist in isolation.
Hence the obvious choice is a splitting into three leptons (the three generations of electrons, represented by $\sigma^1$, $\sigma^2$ and $\sigma^3$) 
and two quarks (the up and down quarks, represented by $\sigma^4$ and $\sigma^5$).

It appears then that the fundamental splitting of fermions into leptons and quarks is therefore already modelled inside the Dirac algebra, and corresponds to a splitting of
$\HH'$ into $\CC+\CC^\perp$. This insight is not at all obvious in the standard formalism, and arises here as a result of the fundamental quaternionic structure
of the Dirac algebra, which is obscured by the standard complex notation. The property of quark confinement is an immediate consequence of their negative mass,
which is another property that is obscured by the factor of $i$ in the mass term in the Dirac equation, which led Dirac to interpret negative mass as negative energy,
and to interpret particles with negative energy as anti-particles. 
In the Jordan algebra, however, anti-particles can possibly be explained by negatives of particles, so that
the mass, obtained by squaring, is the same for both particle and anti-particle.

\subsection{The Lie algebra}
The Lie algebra generated by the anti-Hermitian matrices is $\mathfrak{sp}_2(\HH')$, and 
has a basis consisting of the following matrices:
\begin{align}
&\sigma^1\sigma^2=\begin{pmatrix}0&1\cr -1&0\end{pmatrix},\cr
&\sigma^4\sigma^5=\begin{pmatrix}I&0\cr 0&I\end{pmatrix},\quad
\sigma^2\sigma^3=\begin{pmatrix}I&0\cr 0&-I\end{pmatrix},\quad
\sigma^3\sigma^1=\begin{pmatrix}0&I\cr I&0\end{pmatrix},\cr
&\sigma^3\sigma^5=\begin{pmatrix}J&0\cr 0&J\end{pmatrix},\quad
\sigma^2\sigma^4=\begin{pmatrix}J&0\cr 0&-J\end{pmatrix},\quad 
\sigma^4\sigma^1=\begin{pmatrix}0&J\cr J&0\end{pmatrix},\cr
&\sigma^4\sigma^3= \begin{pmatrix}K&0\cr 0&K\end{pmatrix},\quad
\sigma^2\sigma^5=\begin{pmatrix}K&0\cr 0&-K\end{pmatrix},\quad
\sigma^5\sigma^1=\begin{pmatrix}0&K\cr K&0\end{pmatrix}.
\end{align}
The splitting into $\CC+\CC^\perp$ splits the compact part, $U(2)$, usually interpreted as a spin group $SU(2)$ together with a scalar gauge group $U(1)$,
from two $3$-vector representations of the spin group, that are converted into a single complex $3$-vector by the gauge group.
However, this interpretation begs the question, which is the real $3$-vector that extends $SU(2)$ to the relativistic spin group $SL(2,\CC)$?
Moreover, there is no copy of the gauge group $SU(2)$ for the weak force, that one would expect to see here.
And finally there is an inconsistency with the interpretation of the fermions in the Jordan algebra that was proposed in the previous section.

Moreover, in the Standard Model the spin is always measured in the $z$ direction, and the spins in the $x$ and $y$ directions do not appear.
Similarly, only the third component of weak isospin occurs in the Standard Model. We are therefore not obliged to maintain the symmetry between the $x$, $y$
and $z$ directions of spin, or between the $1$, $2$ and $3$ components of weak isospin.
We can therefore restrict to $\sigma^1\sigma^2$ to represent spin in the $z$ direction, and $\sigma^4\sigma^5$ to represent the third component of weak isospin.
The group $SU(2)$ of symmetries on the indices $1,2,3$ is then available to act on directions \emph{relative} to the direction of spin, rather than absolute directions,
and the group $SL(2,\RR)$ of scalar matrices is available to act on directions of weak isospin \emph{relative} to the third component.

\subsection{Introducing mass}
This allows us to use $SU(2)$ to describe the three generations of electrons, as proposed above, and to describe their different couplings to gravity
(i.e. their different masses) as couplings of the direction of spin to the gravitational field. We then have a choice of perspective, either to fix the direction of spin
and allow the direction of gravity to be arbitrary, or to fix the direction of gravity and allow the direction of spin to be arbitrary.

When we use the
weak interaction to couple neutrinos to electrons, we fix the direction of spin, which is the same as the direction of momentum of the neutrino.
If we then change the direction of the gravitational field, then we change the coupling between gravity and spin, so that the neutrino can interact with
an electron of a different generation. Hence this interpretation provides a basic explanation of neutrino oscillations \cite{oscillation,neutrinos,SNO}
in terms of coupling
of electron spins to gravity. 

As a result of this change of viewpoint, the Lie algebra gives
no global description of $3$-dimensional space, but only a description of $2$-dimensional space relative to 
an arbitrary direction determined by the choice of the observer. We can use the
dynamics of spacetime as described by General Relativity \cite{GR1,GR2,GR3}, 
to pick the third direction as, say, the direction of freefall from the point of phase space under consideration.
In standard approaches to quantum gravity, this is usually done by creating a spin connection over a spacetime manifold, in order
to de-couple gravity from the other forces. However, this seems unnecessarily complicated, if we have the ability to embed the freefall trajectories
directly into phase space.
The flip-side of this approach
is that we are then forced to couple gravity to the other fundamental forces, and calculate or measure the appropriate mixing parameters,
which are essentially the masses of a suitable set of `fundamental' particles.

\section{Adding the strong force}
\subsection{The gauge group and possible interpretations}
To extend the description of quantum dynamics to include the third dimension, 
we must extend the $2\times 2$ matrices to $3\times 3$ matrices,
which extends the fermionic part of the model from $5$ to $14$ degrees of freedom, and extends the bosonic part from $10$ to $21$ degrees of freedom.
The compact part of the Lie algebra now generates a gauge group $U(3)$, which extends the gauge group $U(1)$ of QED by the gauge group $SU(3)$ of the strong force \cite{QCD}.
This group splits the Jordan algebra into $8$ dimensions of symmetric matrices on $1,I$ and $6$ dimensions of anti-symmetric matrices on $J,K$, 
in which we must choose
bases to identify specific particles. 

It is plausible to suppose that the anti-symmetric matrices are the six leptons, but the symmetric matrices cannot be the six quarks, so perhaps they are the baryon octet,
restricting to the proton and neutron in two dimensions. Restriction from $SU(3)$ to $SO(3)$ breaks the symmetry further to $3+3$ leptons and $3+5$ baryons.
This group is both an absolute rotation group in space, and a rotation of colours in quantum chromodynamics (QCD). It therefore does not change the mass of baryons
in QCD, but it does change the gravitational couplings of the leptons. Therefore, as the tidal aspects of the gravitational field vary, the individual masses of the leptons may
change, but the sum of all three is a metric on the Jordan algebra that should remain constant \cite{remarks}. 

\subsection{Weak-strong mixing}
To examine this more closely, note that this group $SO(3)$ commutes with the group $SL(2,\RR)=Sp_1(\HH')$ generated by the scalar matrices, which was previously
identified with the gauge group of the weak interaction. The product of the two is a group $SO(3)\times SL(2,\RR)$,
in which $SO(3)$ acts as the lepton generation symmetry, and $SL(2,\RR)$ acts as the weak gauge group, that separates the proton and neutron. The action of this group
on the $14$ dimensions of fermions is $9+5$, which separates the weak triplets (three negatively charged electrons, three neutrinos and three positively charged protons) from the
 five neutrons, and describes a generation-independent form of beta decay.

Invariance of total energy under rotation of the frame of reference 
implies that the total mass of the $9$ particles must be equal to the total mass of the $5$ particles, in every frame of reference. In our frame of
reference, the neutrino masses are effectively zero, and there is no evidence that they are significant in any other frame of reference either,
which leads to the mass equation
\begin{align}
m(e) + m(\mu) +m(\tau) + 3m(p) = 5m(n),
\end{align}
which is known to hold to well within the limits of experimental uncertainty \cite{remarks}.

\subsection{Fermions}
The $14$ fermions are the traceless Hermitian $3\times 3$ matrices, but it is not completely obvious how to embed the 2-space in the 3-space,
so that the correct basis to use is not obvious. 
The $6$ leptons are anti-symmetric over $J$ and $K$, so in this case the best basis is probably
\begin{align}
&\begin{pmatrix} 0&J&0\cr-J&0&0\cr 0&0&0\end{pmatrix}, \quad
\begin{pmatrix} 0&0&0\cr0&0&J\cr 0&-J&0\end{pmatrix}, \quad
\begin{pmatrix} 0&0&-J\cr0&0&0\cr J&0&0\end{pmatrix}, \cr
&\begin{pmatrix} 0&K&0\cr-K&0&0\cr 0&0&0\end{pmatrix}, \quad
\begin{pmatrix} 0&0&0\cr0&0&K\cr 0&-K&0\end{pmatrix}, \quad
\begin{pmatrix} 0&0&-K\cr0&0&0\cr K&0&0\end{pmatrix}.
\end{align}
Let us for the sake of argument take the first row to represent the three generations of neutrinos (in which momentum is the defining characteristic)
and the second row to represent the three generations of electrons (in which mass is the defining characteristic).

The other $8$ dimensions include $3$ that are antisymmetric over $I$, and $5$ that are symmetric over the real numbers:
\begin{align}
\label{GMmats}
&\begin{pmatrix} 0&I&0\cr-I&0&0\cr 0&0&0\end{pmatrix}, \quad
\begin{pmatrix} 0&0&0\cr0&0&I\cr 0&-I&0\end{pmatrix}, \quad
\begin{pmatrix} 0&0&-I\cr0&0&0\cr I&0&0\end{pmatrix}, \cr
&\begin{pmatrix} 0&1&0\cr1&0&0\cr 0&0&0\end{pmatrix}, \quad
\begin{pmatrix} 0&0&0\cr0&0&1\cr 0&-1&0\end{pmatrix}, \quad
\begin{pmatrix} 0&0&1\cr0&0&0\cr 1&0&0\end{pmatrix},\cr
&\begin{pmatrix} 1&0&0\cr0&-1&0\cr 0&0&0\end{pmatrix},\quad
\begin{pmatrix} 0&0&0\cr0&1&0\cr 0&0&-1\end{pmatrix}.
\end{align}
These matrices are essentially the same as the Gell-Mann matrices, but here they represent fermions, not gluons.
The basis for the diagonal matrices is not obvious, but if this is the baryon octet, with rows and columns corresponding to up, down and strange quarks,
then the diagonal matrices represent the $\Lambda$ and $\Sigma^0$ baryons,
both made of one of each 
quark, and the mixing between these is quite complicated in the Standard Model.
The other two triplets are the two triplets that have the same total mass according to the Coleman--Glashow relation, forming the two rows of the display.
The first entry in each row is made of first-generation quarks, and
it would be reasonable to suppose that the real matrix is the neutron, and the imaginary matrix is the proton.
The Coleman--Glashow relation then follows from the invariance of internal energy under rotation of the frame of reference.

\subsection{Anti-Hermitian matrices}
The full gauge group is $Sp_3(\HH')\cong Sp_6(\RR)$ with $21$ degrees of freedom. The electro-strong part of this is the compact subgroup $U(3)$, which mixes with the weak
$Sp_1(\HH')$ to generate the remaining $10$ degrees of freedom. The generators for $SU(3)$ are the anti-Hermitian versions of the Gell-Mann matrices \cite{GellMann},
 i.e. the matrices in
(\ref{GMmats}) multiplied by $I$ (or $-I$), and are extended to $U(3)$ by the scalar $I$. The latter matrix is also in the weak gauge group $SL(2,\RR)$.
The remaining $10$ dimensions consist of $5$ symmetric matrices each over $J$ and $K$, and extend the subgroup $SO(3)$ of $SU(3)$ to two different
groups $SL(3,\RR)$, each of which extends to $GL(3,\RR)$ by adjoining the corresponding 
scalar matrix. 

\begin{align}
&\begin{pmatrix} 0&J&0\cr J&0&0\cr 0&0&0\end{pmatrix}, 
\begin{pmatrix} 0&0&0\cr0&0&J\cr 0&J&0\end{pmatrix}, 
\begin{pmatrix} 0&0&J\cr0&0&0\cr J&0&0\end{pmatrix},\cr
&\begin{pmatrix} J&0&0\cr0&-J&0\cr 0&0&0\end{pmatrix},
\begin{pmatrix} 0&0&0\cr0&J&0\cr 0&0&-J\end{pmatrix},\cr
&\begin{pmatrix} 0&K&0\cr K&0&0\cr 0&0&0\end{pmatrix}, 
\begin{pmatrix} 0&0&0\cr0&0&K\cr 0&K&0\end{pmatrix}, 
\begin{pmatrix} 0&0&K\cr0&0&0\cr K&0&0\end{pmatrix},\cr
&\begin{pmatrix} K&0&0\cr0&-K&0\cr 0&0&0\end{pmatrix},
\begin{pmatrix} 0&0&0\cr0&K&0\cr 0&0&-K\end{pmatrix}.
\end{align}

\subsection{Planck's constant}
Quantisation arises from the fact that $I^2=-1$, so that $e^{2\pi I}=1$ and one can count the number of rotations around the circle. 
The counting is implemeted by introducing Planck's constant $h$
as the circumference of the circle. 
Since Planck's constant has units of angular momentum, that is momentum times distance, it makes sense to assign momentum to $J$ and distance to $K$ (or
vice versa). Then Planck's constant has a second interpretation as the \emph{area} enclosed by the circle in phase space.
The anti-commutation of $J$ and $K$ 
enforces the basic property of angular momentum that if the momentum and position coordinates are interchanged, then the angular momentum is negated.
In $3$-dimensional space this is the fact that the angular momentum is defined as $\mathbf p\times \mathbf q=-\mathbf q\times \mathbf p$, 
where $\mathbf p$ and $\mathbf q$ are the
momentum and position vectors in phase space relative to the chosen origin of coordinates.
 The quantisation then arises automatically from the 
 fact that $J\mathbf p\times K\mathbf q = I (\mathbf q\times \mathbf p)$, 
 without the need to quantise space or momentum separately. 
 
 In other words, our quaternionic notation for the electro-weak and strong forces in particle physics translates
directly to classical Hamiltonian mechanics, without the need for any complexification or other tweaks.
The Dirac algebra that is supposed to incorporate the Lorentz group in the form $SL(2,\CC)=Spin(3,1)$, mixing all three dimensions of space and one of time,
appears in fact to mix four dimensions of \emph{phase space} instead, and only two dimensions of space and one of time. Moreover, the mixing of space with time
is not necessary in order to reproduce the entire algebraic structure of the Standard Model.
What is necessary instead is a mixing of position with momentum, so that two observers who measure different position and momentum coordinates for
an event can nevertheless agree about the physical process that they have measured, assuming that they use Hamiltonian mechanics to analyse the
dynamics of the physical situation.

\subsection{Observers}
The main point of this remark is that the two observers only need to agree on the transformation between phase space coordinates. 
They can then calibrate their measurements of Planck's constant, and hence of energy and time and the speed of light, but there is no need for
them to agree about any properties of mass at all. They can both use whatever definitions of mass they prefer, and still describe and predict the same physical events.
For example, they may find it useful to agree on a standard value for the mass ratio of electron and proton, for practical purposes, but this is entirely unnecessary for
the prediction of properties of fundamental physical processes.

By this means, the right-handed part of the Dirac spinor is directly identified with an element of phase space in two dimensions, in such a way that the third dimension
represents the direction of the angular momentum of the chosen frame of reference relative to the local inertial frame, defined by freefall within the gravitational field.
The extension of the Dirac spinor to three dimensions therefore allows us to include gravity within the general quantisation scheme.

\section{Gravity}
\subsection{Principles of Hamiltonian dynamics}
Hamiltonian mechanics is based on four fundamental notions: space or position $\mathbf q$, time $t$, momentum $\mathbf p$ and energy $H$ (the Hamiltonian). 
Hamilton's equations are
\begin{align}
\label{Heq}
\frac{d\mathbf q}{dt} = \frac{\partial H}{\partial \mathbf p}, \quad \frac{d\mathbf p}{dt} = - \frac{\partial H}{\partial \mathbf q}.
\end{align}
The left-hand side of the first equation is the velocity, and the left-hand side of the second equation is the force.
Together they imply conservation of energy, since
\begin{align}
\frac{dH}{dt} & = \frac{\partial H}{\partial\mathbf p}.\frac{d\mathbf p}{dt} +\frac{\partial H}{\partial\mathbf q}.\frac{d\mathbf q}{dt}\cr
& = \frac{d\mathbf q}{dt}. \frac{d\mathbf p}{dt} - \frac{d\mathbf p}{dt} \frac{d\mathbf q}{dt}\cr 
& = 0. 
\end{align}
The great advantage of the Hamiltonian approach is that it is possible to calculate the dynamics of a system without knowing the masses of anything in the system.
Therefore Hamiltonian mechanics is routinely used in most branches of physics, especially in quantum mechanics and condensed matter physics.
But even more important than that,
there is a very obvious duality between $\mathbf p$ and $\mathbf q$, such that if you switch the two, and negate one of them, the equations remain the same.

One can also mix and match the two by taking linear combinations of the equations, and you can mix the three coordinates of momentum provided you mix the
coordinates of space in the dual manner. A duality of this kind is a symplectic duality, so that the group of all coordinate transformations of phase space that preserve
the Hamilton equations is a symplectic group, so can be written in terms of quaternions. If we want to interpret momentum and position as real vectors, then they must go in the
$J$ and $K$ coordinates of the split quaternions, so that multiplication by $I$ converts position into momentum and momentum back to the opposite position, in exactly the way that a spinning object in two dimensions behaves.
Hence the group of coordinate transformations between possible observers in 2-dimensional dynamics is $Sp_2(\HH')$, acting on phase space in exactly the same way that
the Dirac algebra, or its subgroup $Spin(3,2)$, acts on spinors.

\subsection{Principles of relativity}
In order to describe gravity in 3-dimensional dynamics, we must extend the gauge group from $Sp_2(\HH')$ to $Sp_3(\HH')$ to parametrise all possible changes of
phase space coordinates between different observers. The General Principle of Relativity (GPR)  is the principle that all observers observe the same physics,
independently of their choices of phase-space coordinates. Notice, however, that that is not the same as the Principle of General Covariance (PGC), that is the principle
that observers can use arbitrary \emph{spacetime} coordinates. The latter principle has a gauge group $GL(4,\RR)$, which is not the same as
 $Sp_3(\HH')$. The fact that General Relativity (GR) is based on PGC rather than GPR \cite{GR1,GR2,GR3}
 means that GR may not in fact satisfy the GPR on extreme scales where it has not been
 thoroughly tested \cite{Rubin,Chae}. 
 
 It may be possible to embed both groups into a larger group such as $Sp_4(\HH')$ in order to reconcile their differences \cite{PGM}, or it may be possible to
 modify GR to use the gauge group $Sp_3(\HH')$, without affecting calculations on a small timescale. The latter proposal is considerably simpler,
 and therefore we should look at this one first. The gauge group $Sp_3(\HH')$ can only be used if there is a universal timescale, but it is independent of length scale,
 and should therefore be able to provide a model of gravity that can be applied on all scales. 
 
 What particularly distinguishes this gauge group from GR is that it allows for transformations between
 rotating frames of reference on arbitrary scales, and therefore explicitly incorporates Mach's Principle,
 that rotation with respect to the large-scale universe can be felt locally, and that the equations of physics are simpler in an inertial frame.

 Although the universe has no physical centre at all, we have to 
 put the mathematical centre somewhere, and Einstein says we can put it wherever we like. Mach's Principle, however, tells us that we can \emph{detect} 
 our rotations and accelerations in the wider universe by measuring \emph{mass}.
 More or less any mass measurement will do,
 for example measuring the mass ratios of elementary particles,
 or mass ratios of different copies of the International Prototype Kilogram (IPK).

\subsection{Algebraic structure of the tensors}
By working now over the algebra of split quaternions $\HH'$, we have Hermitian and anti-Hermitian tensors written as $1\times 1$, $2\times 2$ and $3\times3$ matrices.
The $3\times 3$ case contains $15$ Hermitian tensors, forming a Jordan algebra splitting as $1+14$, and $21$ anti-Hermitian tensors,
forming an irreducible Lie algebra and generating the gauge group. This compares to the GR gauge group $SL(4,\RR)=Spin(3,3)$, whose irreducible $15$-dimensional
Lie algebra consistis of $6\times 6$ anti-symmetric
real matrices, acting on the Jordan algebra of symmetric matrices, which splits as $1+20$ to give the Riemann Curvature Tensor.
The two things look more or less the same, except for the fundamental difference that the Jordan algebra and the Lie algebra have been swapped round.

If we now look at the $2\times 2$ case, then in the symplectic Hamiltonian model the Jordan algebra splits as $1+5$ and the Lie algebra is irreducible of dimension $10$.
Again, we see in GR the tensors on real $4$-space splitting as the Lie algebra for $SO(3,1)$ of dimension $6$, plus the Jordan algebra of symmetric tensors,
representing the Ricci tensor, which splits as $1+9$ into the Ricci scalar plus the Einstein tensor.
In the $1\times 1$ case, the Hamiltonian model contains a trivial $1$-dimensional Jordan algebra, plus a $3$-dimensional Lie algebra of the gauge group $SL(2,\RR)$,
while the GR model on $2\times 2$ real matrices contains a $1$-dimensional Lie algebra, generating a gauge group $U(1)$, plus a $3$-dimensional Jordan algebra,
splitting as $1+2$.

\subsection{The source of the problem}
The incompatibility between GR and particle physics now looks to be due to a systematic swapping of the Lie algebra with the Jordan algebra between the two theories.
If we suppose that one is right and the other is wrong, then which one is it? We have already seen that the SMPP is compatible with the basic principles of Hamiltonian dynamics,
provided we interpret the unexplained concepts such as spinors and colours appropriately in terms of phase space. The gauge groups of particle physics embed suitably
in the overall gauge group $Sp_3(\HH')$, but they do not all embed in the gauge group $SL(4,\RR)$ used in GR. Moreover, the latter gauge group is incompatible with the basis
principles of Hamiltonian dynamics, since it does not embed in $Sp_3(\HH')$. Therefore we must conclude that SMPP is essentially correct, as is strongly confirmed by experiment,
and GR is incorrect, and does not scale from Solar System dynamics to galaxy dynamics, as is 
strongly confirmed by astronomical observations \cite{Rubin,Chae,Hernandez,newparadigm}.

The mathematical inconsistency at the heart of this confusion is to put an orthogonal duality on to phase space, gauged by $SO(3,3)$, when the natural duality on phase space is
in fact symplectic \cite{deGosson}, and is gauged by $Sp_3(\HH')\cong Sp_6(\RR)$. There is therefore no possible way to re-interpret GR as a correct Hamiltonian theory of gravity.
However, it works on a small scale, in which the timescales are small, and there is a fixed scale of angular momentum, measured in units of Planck's constant,
with which to ensure that the Hamiltonian duality between position and momentum is approximately satisfied.

An equivalent way to look at this distinction is to look at Lorentz transformations as transformations on phase space, rather than on spacetime. Lorentz transformations
on $1+1$ spacetime generate $SO(1,1)$, while on $1+1$ phase space they generate $GL(1,\RR)\cong Z_2\times SO(1,1)$. Hence the standard interpretation works fine
by just ignoring the sign. In $2+2$ phase space we have $GL(2,\RR)$, while on $2+1$ spacetime we get $SO(2,1) \cong GL(2,\RR)/GL(1,\RR)$, so again we can recover the
standard interpretation by ignoring the real scalars. But in $3+3$ phase space the group $GL(3,\RR)$ is completely unrelated to the Lorentz group $SO(3,1)$.

\subsection{Quantum gravity}
With these preliminaries, we have a clear strategy for constructing a quantum theory of gravity. Moreover, this quantum theory of gravity already exists: it is called the
Standard Model of Particle Physics. All we need to do is adjust the interpretations a little bit.
The main issue is that by swapping Lie algebras with Jordan algebras, we have inadvertently swapped fermions with bosons. All attempts to add such a `supersymmetry'
between fermions and bosons have failed experimentally. They also fail theoretically, because the dimension of the Lie algebra is different from
the dimension of the Jordan algebra.

Therefore, the theoretical spin 2 graviton proposed by GR is not a boson at all, it is a fermion. It exists in the SMPP, and is called the neutrino. Normally, one would 
assume that this Jordan algebra consists of three neutrinos and three anti-neutrinos, one each for each generation of electrons. But the Jordan algebra splits as $1+5$,
so there are only $5$ dimensions of neutrinos, not $6$. The neutrinos therefore oscillate between flavours under the influence of the tidal forces of gravity.
The scalar that splits off is the energy, which for a given observer in a particular gravitational field can also be interpreted as mass. But mass is a Newtonian concept,
or strictly speaking two Newtonian concepts (inertia and gravity),
that does not exist in true Hamiltonian dynamics, so we are better off without it \cite{universal}.

\section{Conclusion}
By reducing the SMPP and GR to their fundamental algebraic constituents, we have traced the inconsistency between them to a switch between 
Lie algebras and Jordan algebras. Since there is no mathematical equivalence between these two types of algebras, there is no way to reconcile the
two theories except by declaring that (at least) one of them is wrong. By applying the basic principles of Hamiltonian dynamics and gauge theory,
we have found no flaw in SMPP, beyond a few details of interpretation. This conclusion is also strongly supported by experiment.

On the other hand, we have found that GR is inconsistent with the fundamental principle
that every observer must be able to describe physical reality in their own coordinate system for phase space.
We are therefore obliged to conclude that GR is fundamentally flawed, and must be abandoned as a theory of gravity.
It gives correct answers on a Solar System scale in which only small perturbations from Newtonian gravity are required,
but it is not scale-invariant (renormalizable), so cannot be applied on significantly larger scales.

Astronomical observations have by now demonstrated that the scale at which Newtonian gravity for isolated binary star
systems fails
in the local part of the Milky Way
is approximately three orders of magnitude greater than the Earth's orbit around the Sun \cite{Chae,Hernandez}. 
This is approximately the same as the limit of the Sun's influence over the 
Solar System, at the point where the gravitational pull of the Sun starts to merge into the background of the rest of the galaxy. 

This is the point, therefore, at which the modelling of the orbits of the two stars around each other changes from being a two-dimensional problem
in which the stars can be treated in isolation from the rest of the universe, to a three-dimensional problem in which the centre of the galaxy
must also be included in the modelling data. The orbit of the combined binary star system around the galaxy occupies one quaternionic dimension in
phase space, and the orbits of the two stars around each other occupy the other two. The dynamics are therefore completely different from the dynamics of
a single orbiting system. Moreover, the angle of inclination of the binary star system to the galactic plane is a crucial parameter in these dynamics.
Since Hamiltonian dynamics is renormalizable, such angles of inclination should also appear in particle physics, if a calibration of inertial mass against 
gravitational mass is required \cite{universal,1969,Einstein1919}.

Of course, the suggestions made in this paper do not amount to a new model of physics. The purpose of the paper is solely to identify and eradicate the mathematical inconsistencies
that are known to exist in the standard models, so as to provide a mathematically consistent and rigorous foundation on which it should be possible to build
a revised model, that differs from the accepted standard ones in only a few small but important details.


\begin{thebibliography}{99}

\bibitem{Dirac} P. A. M. Dirac (1928), The quantum theory of the electron,
{\it Proc. Roy. Soc A} {\bf 117}, 610--624.


\bibitem{BjorkenDrell}
J. D. Bjorken and S. D. Drell (1964),
{\it Relativistic quantum mechanics and relativistic quantum fields},
McGraw-Hill.


 \bibitem{Griffiths} D. Griffiths (2008), {\it Introduction to elementary particles},
2nd ed, Wiley.



\bibitem{WoitQFT} P. Woit (2017), {\it Quantum theory, groups and representations}, Springer.

 \bibitem{Zee}
 A. Zee (2016), {\it Group theory in a nutshell for physicists},
 Princeton University Press.


\bibitem{oscillation}
B. Pontecorvo (1968),
Neutrino experiments and the problem of conservation of leptonic charge,
{\it Soviet Phys. JETP} {\bf 26}, 984--988.


\bibitem{neutrinos}
V. Gribov and B. Pontecorvo (1969),
Neutrino astronomy and lepton charge,
{\it Phys. Rev. D} {\bf 22} (9), 2227--2235.

\bibitem{SNO} A. Bellerive et al. (SNO collaboration) (2016),
The Sudbury neutrino observatory, {\it Nuclear Phys. B} {\bf 908}, 30--51. arXiv:1602.02469.

\bibitem{GR1} A. Einstein (1916), Die Grundlage der allgemeinen
Relativit\"atstheorie, {\it Annalen der Physik} {\bf 49} (7), 769--822.


 \bibitem{GR2} A. Einstein (1955),
{\it The meaning of relativity}, 5th ed., Princeton UP.



 \bibitem{GR3}
 G. 't Hooft (2001),
 {\it Introduction to general relativity},
 Rinton.



\bibitem{QCD} W. Greiner, S. Schramm and E. Stein (2007), {\it Quantum Chromodynamics}, Springer.

\bibitem{remarks} R. A. Wilson (2022),
Remarks on the group-theoretical foundations of particle physics,
{\it Intern. J. of Geom. Methods in Modern Phys.} {\bf 19}, 2250164.


\bibitem{GellMann} M. Gell-Mann (1961), The eightfold way: a theory of strong interaction symmetry,
Synchrotron Lab. Report CTSL-20, Cal. Tech.


\bibitem{Rubin} V. Rubin and W. K. Ford, Jr. (1970), 
Rotation of the Andromeda Nebula from a spectroscopic survey of emission regions,
{\it Astrophysical J.} {\bf 159}, 379.


\bibitem{Chae} K.-H. Chae (2023), Breakdown of the Newton--Einstein standard gravity at low accelerations in 
internal dynamics of wide binary stars,
{\it The Astrophysical Journal} {\bf 952}, 128.

\bibitem{Hernandez} X. Hernandez (2023), {\it MNRAS} {\bf 525}, 1401.

\bibitem{PGM} R. A. WIlson (2024), A discrete model for Gell-Mann matrices, arXiv:2401.13000.

\bibitem{newparadigm} P. Kroupa, M. Pawlowski and M. Milgrom (2012),
The failures of the standard model of cosmology require a new paradigm,
{\it Intern. J. Modern Phys. D} {\bf 21} (14).
arXiv:1301.3907.

\bibitem{deGosson} M.. de Gosson and B. Hiley (2011), Imprints of the quantum world in classical mechanics,
{\it Found. Phys.} {\bf 41}, 1415-1436.

\bibitem{universal} R. A. Wilson (2022), Is there a universal concept of mass in fundamental physics? arXiv:2205.05443.


\bibitem{1969} B. N. Taylor, W. H. Parker and D. N. Langenberg (1969), 
Determination of $e/h$, using macroscopic quantum phase coherence in superconductors:
implications for quantum electrodynamics and the fundamental physical constants,
{\it Reviews of modern physics} {\bf 41}(3), 375--496.


\bibitem{Einstein1919} 
A. Einstein (1919), Spielen Gravitationsfelder im Aufbau der materiellen Elementarteilchen
eine wesentliche Rolle? {\it Sitzungsberichte der Preussisschen Akad. d. Wissenschaften}.




\end{thebibliography}
\end{document}